\documentclass{pasj00}
\draft

\def\simlt{\lower.5ex\hbox{$\; \buildrel < \over \sim \;$}}
\def\simgt{\lower.5ex\hbox{$\; \buildrel > \over \sim \;$}}
\def\cm3{{\rm cm^{-3}}}
\def\kms{km s$^{-1}$}

\def\boone{B0104$-$72.3}
\def\eoone{B0104$-$72.3}
\def\akari{{\em AKARI}}
\def\tex{T_{\rm ex}}
\def\nh2{N({\rm H}_2)}
\def\um{$\mu$m}

\begin{document}
\SetRunningHead{Author(s) in page-head}{Running Head}
\Received{2007/2/}
\Accepted{2007/3/}

\title{{\em AKARI} Detection of the Infrared-Bright Supernova Remnant \boone\ in the 
Small Magellanic Cloud}


\author{Bon-Chul \textsc{Koo}\altaffilmark{1}
Ho-Gyu \textsc{Lee}\altaffilmark{1}
Dae-Sik \textsc{Moon}\altaffilmark{2}
Jae-Joon \textsc{Lee}\altaffilmark{1}
Ji Yeon \textsc{Seok}\altaffilmark{1}
Hyung Mok \textsc{Lee}\altaffilmark{1}
Seung Soo \textsc{Hong}\altaffilmark{1}
Myung Gyoon \textsc{Lee}\altaffilmark{1}
Hidehiro \textsc{Kaneda}\altaffilmark{3}
Yoshifusa \textsc{Ita}\altaffilmark{3}
Woong-Seob \textsc{Jeong}\altaffilmark{3}
Takashi \textsc{Onaka}\altaffilmark{4}
Itsuki \textsc{Sakon}\altaffilmark{4}
Takao \textsc{Nakagawa}\altaffilmark{3}
Hiroshi \textsc{Murakami}\altaffilmark{3}
}
\altaffiltext{1}{Department of Physics and Astronomy, Seoul National University, Seoul
151-742, KOREA}
\email{koo@astrohi.snu.ac.kr; hglee@astro.snu.ac.kr; jjlee@astro.snu.ac.kr;
jyseok@astro.snu.ac.kr; hmlee@astro.snu.ac.kr; sshong@astro.snu.ac.kr;
mglee@astro.snu.ac.kr}
\altaffiltext{2}{Department of Astronomy and Astrophysics, University of Toronto, 
Toronto, ON M5S 3H4, CANADA}
\email{moon@astro.utoronto.ca}
\altaffiltext{3}{Institute of Space and Astronautical Science, Japan Aerospace
Exploration Agency, Sagamihara, Kanagawa 229-8510, JAPAN}
\email{kaneda@ir.isas.jaxa.jp; yita@ir.isas.jaxa.jp; jeongws@ir.isas.jaxa.jp; 
nakagawa@ir.isas.jaxa.jp; hmurakam@ir.isas.jaxa.jp}
\altaffiltext{4}{Department of Astronomy, Graduate School of Science, 
University of Tokyo, Bunkyo-ku, Tokyo 113-0003, JAPAN}
\email{onaka@astron.s.u-tokyo.ac.jp; isakon@maitta.astron.s.u-tokyo.ac.jp}

%


\maketitle

\begin{abstract}

We present a serendipitous detection of the infrared-bright 
supernova remnant (SNR) \boone\ in the Small Magellanic Cloud 
by the Infrared Camera (IRC) onboard \akari. 
An elongated, partially complete shell 
is detected in all four observed IRC bands covering 2.6--15 \um. 
The infrared shell surrounds radio, optical, and X-ray emission 
associated with the SNR and is probably a radiative SNR shell. 
This is the {\em first} detection of a SNR shell in
this near/mid-infrared waveband in the Small Magellanic Cloud.  
The IRC color indicates that the infrared emission might be 
from shocked H$_2$ molecules with some possible contributions from 
ionic lines. We conclude that \boone\ is a middle-aged SNR 
interacting with molecular clouds, similar to the 
Galactic SNR IC 443. 
Our results highlight the potential of \akari\ IRC observations
in studying SNRs, especially for diagnosing SNR shocks.
 
\end{abstract}

\section{Introduction}

The Small Magellanic Cloud (SMC) has about twenty known SNRs. 
They have been identified and studied 
mostly in radio, X-ray, and optical bands 
(\cite{mat84, hey04, fil05} and references therein).
Infrared studies are very limited in spite of 
their advantages in understanding the interstellar medium of 
this nearby, metal-poor irregular galaxy
as well as in understanding the nature of SNRs.  

In this paper, we report the detection of the SNR 
\boone\ (J010619$-$720527; IKT 25; hereafter \eoone) in the SMC 
in the 2.6--15 \um\ range by the Infrared Camera (IRC) onboard \akari.
This is the {\em first} SNR detected in this IR band in the SMC. 
The only other SNR detected in mid-infrared waveband 
in the SMC is the radio/X-ray bright SNR B0102.2$-$72.19, which was 
observed by \citet{sta05} using the
{\it Spitzer Space Telescope} and detected only at 24 \um.
\eoone\ is the faintest radio SNR in the SMC, 
located in the northwestern area of the ``wing'' close to the ``bar''.  
It is small ($100''$ or 29 pc at the distance of 60 kpc), and its radio structure is 
not well resolved \citep{mat84, fil05}. 
The faintness together with its relatively small spatial extent 
led \citet{mat84} to suggest that it is possibly a 
Balmer-dominated Type Ia SNR. 
On the other hand, the remnant has a flat 
radio spectrum with a spectral index 
$\alpha=0.19\pm 0.28$ ($S_\nu\propto \nu^\alpha$; \cite{fil05}), 
which suggests a possibility of Crab-like SNR or `plerion'.
The remnant has been studied in X-rays using {\it ROSAT} and {\it XMM}. 
\citet{hug94} 
showed that the X-ray emission is composed of a
diffuse emission extending along the north-south direction and 
a point-like source at the southern end. 
They tentatively identified the point-like source as a 
Be/X-ray binary system and proposed a core-collapse SN origin for 
the remnant. \citet{hey04} obtained an X-ray spectrum of the remnant 
using the {\it XMM-Newton} observatory and showed that the 
Fe abundance is enhanced 
with respect to the mean O, Ne and Mg abundance values.
They concluded that the X-ray emission is possibly from ejecta remains of 
Type Ia. In summary, the radio and X-ray observations do not 
yield a coherent picture on the nature of the remnant.
We show that \eoone\ has a well defined 
IR shell surrounding the radio, optical, and X-ray emissions and that 
\eoone\ is probably interacting with molecular clouds. We discuss 
the nature of the SNR based on \akari\ observations.

\section{Serendipitous Detection}

\eoone\ was detected serendipitously 
during the pointed observation of
the oxygen-rich SNR B0103$-$72.6 (DEM S125; \cite{park03}) in the SMC 
using the \akari\ IRC on November 8, 2006.
The IRC is equipped with three waveband channels 
with 10$'$ $\times$ 10$'$ field-of-view (FOV),
which operate simultaneously at any exposure time (\cite{ona07}).
Among them the NIR and MIR-S channels share the same FOV, 
while the MIR-L channel has FOV 25$'$ away from them.
\eoone\ happened to be in the NIR/MIR-S FOV  
when B0103$-$72.6 was centered at the MIR-L FOV.
We used the two-filter mode (IRC02), which 
gave two band images at each channel:
N3 and N4 in NIR, and S7 and S11 in MIR-S.
N3, N4, S7, and S11 images have isophotal wavelengths of 
3.19, 4.33, 7.12 and 10.45 $\mu$m with effective bandwidths of 
0.87, 1.53, 1.75 and 4.12 $\mu$m, respectively.
The pixel sizes of the NIR and MIR-S channels are 
$1.''46$ and $2.''34$, respectively.
The angular resolutions are 
$4.''0$, $4.''2$, $5.''1$, and $4.''8$ 
in N3, N4, S7, and S11, respectively \citep{ona07}.
The total on-source integration times were
178 s for NIR and 196 s for MIR-S.
The basic calibration and data handling 
including dark subtraction, linearity fitting, distortion correction,
flat fielding, and image combining were processed using the 
standard IRC Imaging Data Reduction Pipeline version 070104
\citep{ita07}. 
We, however, used our own flat images derived from 
the North Ecliptic Pole data around our observing date for the MIR-S,   
because the IRC MIR-S flats have noticeable ``dark pattern''
and, if we use the provided superflats, it 
leaves an artifact even after the flat-fielding (see IRC Data User's Manual).
The astrometric solutions were obtained by matching with 
the coordinates of 2MASS stars in the field using 
the Pipeline process {\it putwcs}.
The resulting positional uncertainty ($1\sigma$) is $\le 0.''4$ for 
N3, N4, S7, and $1''$ for S11.

\begin{figure}
  \begin{center}
    \FigureFile(110mm,110mm){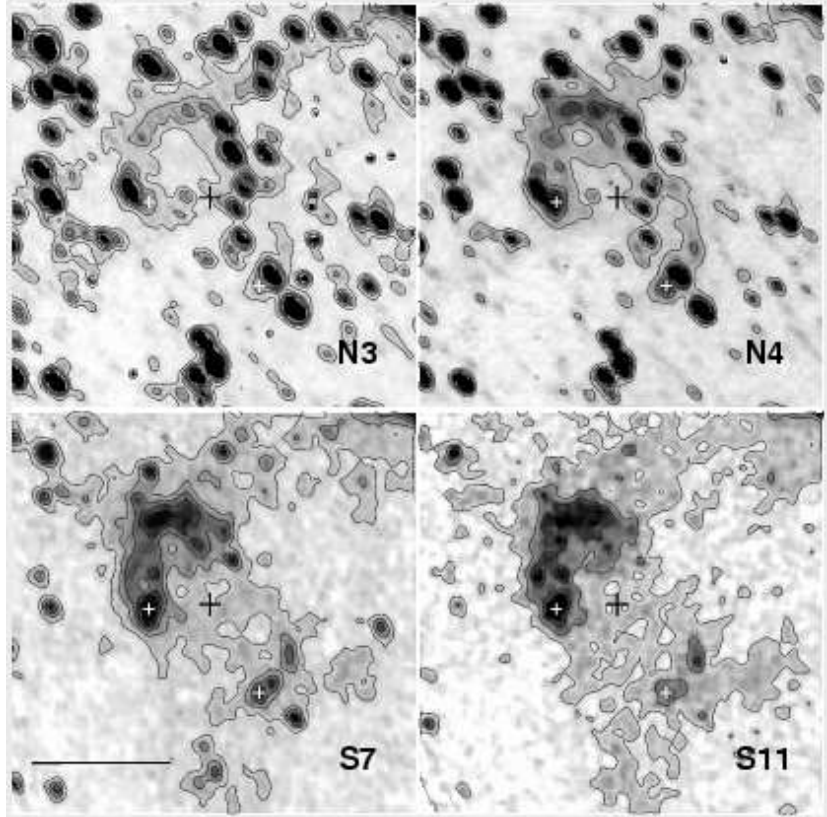}
  \end{center}
  \caption{\akari\ images of the SNR \boone\ in N3, N4, S7, and S11 bands. 
Contour levels are linearly spaced
starting at 0.18, 0.30, 3.90, and 16.4 MJy sr$^{-1}$ 
with steps of 0.04, 0.04, 0.06, and 0.1 MJy sr$^{-1}$ in 
N3, N4, S7, and S11 images, respectively.
The N3 and N4 images show trails of pixels with enhanced brightness, 
which is a detector artifact called 
multiplexer bleed, or ``muxbleed'' (IRC Data User's Manual).
The black cross marks the geometrical center of the shell 
at (01$^{\rm h}$ 06$^{\rm m}$ 18.$^{\rm s}$3,$-72^\circ$ 05$'$ 37$''$). 
The white crosses mark the peak positions in the S7 band.
The scale bar in the S7 image represents an angular size of $1'$.
North is up and east is to the left.}\label{fig:fig1}
\end{figure}

Fig. 1 shows \akari\ images of \eoone. 
A partially-complete shell structure is apparent in all four bands.
The shell is elongated along the northeast-southwest direction and 
has a bright, well-defined circular 
northeastern portion (hereafter the IR-NE shell).
The brightness of the IR-NE shell is not uniform, and there is a 
prominent bright spot at the southern end which is 
marked by a white cross. 
The maximum, background-subtracted brightness of 
the spot is $\sim 0.11$, 0.17, 0.41, and 0.42 MJy sr$^{-1}$ in 
N3, N4, S7, and S11, respectively.
The rest part of the shell
is mostly missing, but, to the southwest,  
a small segment of the shell (the IR-SW shell), which 
has a curvature opposite to the IR-NE shell, is visible. 
Some faint, filamentary emission is also visible 
between the two shell parts.
These IR-NE and IR-SW shells 
surround the radio emission neatly (see next section),
so that they are likely to be parts of the SNR shell.
The extent of the SNR determined from Fig. 1 is 
$100''\times 65''$ (or 29 pc $\times$ 19 pc).

\section{Comparison with Other Waveband Images}

\begin{figure}
  \begin{center}
    \FigureFile(110mm,110mm){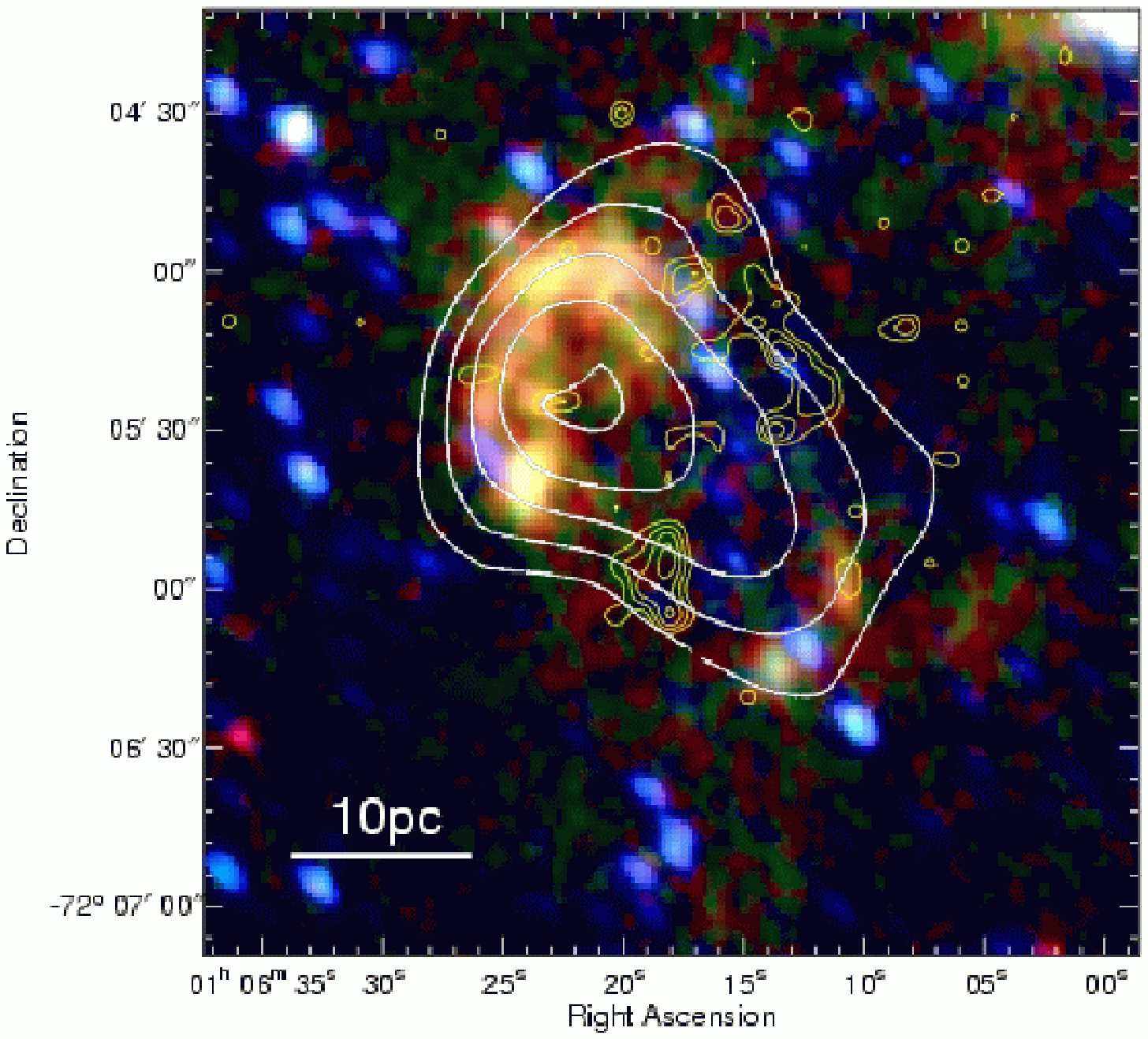}
  \end{center}
  \caption{\akari\ three-color image generated from N4 (B), S7 (G), and S11 (R).
White contours show the 843 MHz radio map of \citet{mat84} while 
yellow contours show the {\it ROSAT} HRI 0.5--2.0 keV map. 
}\label{fig:fig2}
\end{figure}

Fig. 2 is our three-color image representing the N4 (B), S7 (G), and S11 (R) 
emission of \eoone. The white contours represent the 
843 MHz radio map (FWHM=$43''$) 
from the Molonglo Observatory Synthesis Telescope 
\citep{mat84}. The radio emission is elongated along the 
northeast-southwest direction and almost perfectly 
fits into the IR shell. 
It has the maximum inside the IR-NE shell and 
an extended `plateau' to southwest, which is confined 
to a relatively narrow region.
The 2.37 GHz map (FWHM=$40''$) of \citet{fil05} obtained by using 
the Australian Telescope Compact Array shows essentially 
the same radio morphology.  
It is possible that the remnant has a shell structure in the radio too but 
is smeared because of the poor resolutions of 
these interferometric observations.
We have simulated how the remnant would appear with a large beam 
if the radio brightness is proportional to that of the IR shell. 
We convolved the S7 image, where the shell is most prominent, 
with a Gaussian beam of FWHM=$43''$. 
The simulated map also 
shows a centrally brightened remnant, but its 
peak occurs close to the IR-NE shell,
$\sim 15''$ north to the peak of the 843 MHz emission. 
Another difference between the simulated and the observed radio maps
is that, in the former, there is a separate peak at the position 
of the IR-SW shell while, in the latter, the intensity decreases 
smoothly toward the southwest. 
The above comparison leads us to conclude that the radio 
brightness distribution should be 
different from the IR brightness distribution. 
The radio emission 
may peak inside the IR-NE shell and have some 
extended emission toward the southwest. 

The yellow contours in Fig. 2 are 
the 0.5--2.0 keV {\it ROSAT} High-Resolution Imager (HRI) data provided 
by {\it SkyView} 
and show that X-rays are distributed across the central part of the remnant 
along the southeast-northwest direction (see also Fig. 3 of \cite{hug94}). 
There is no apparent X-ray emission associated with the IR shell.  
Toward the northwest, 
the X-ray emission extends beyond the boundary connecting the 
IR NE and SW shells. 
The bright, compact X-ray source near the southeastern 
boundary of the remnant is the source 
identified as a Be/X-ray binary system by \citet{hug94}. 
\citet{hey04} showed that the X-ray spectrum of the 
diffuse emission can be fitted by a 
thermal plasma model at 0.6 keV. 

In the optical, \eoone\ has a complex morphology \citep{mat84}. 
We can see optical filaments coincident with the IR shells in 
the H$\alpha$ and [S II] $\lambda\lambda$ 6717/6731 images 
of \citet{mat84}. 
The bright spot in the IR-NE shell appears bright in the optical 
too, but the rest of the shell is generally faint. 
Instead there is brighter, filamentary optical 
emission all over the southwestern area 
and also in the northeastern area of the remnant. 
These optical emission features seem to correspond to the 
faint IR filamentary features in the N4/S7 images of Fig. 1, 
although the comparison is limited by the resolution and sensitivity.
(We can also identify some corresponding 
filaments in their [O III] $\lambda$ 5007 image but 
only tentatively due to poor S/N ratio of the image.) 
In short, the IR shells have associated optical emission 
but this emission is 
faint. Instead there are bright optical filaments associated
with faint IR features. 

\section{IRC Colors and the Origin of IR Emission}

In this section, we derive IRC colors of the 
IR shells and explore the origin of the IR emission. 
We derive the IRC fluxes of the shells 
by summing all signals inside an ellipse surrounding 
each of them.  
Point sources are excluded by putting masks 
on them. The same areas are masked out in all bands even if 
they are visible only in the NIR bands.
The background 
contribution is estimated from surrounding annulus and is subtracted. 
The contamination due to the muxbleed in the bands N3 and N4
is also estimated and subtracted.
The conversion to astronomical fluxes is performed by 
using the conversion factors in the IRC Data Users Manual,
which were derived from standard stars. 
The fluxes are not extinction-corrected because the extinction is small.
\citet{hey04} obtained an H nuclei column density of
$4.8\pm 1.7 \times 10^{21}$~cm$^{-2}$ toward the source,
while the Galactic hydrogen column density toward the SMC is
$6.1 \times 10^{20}$~cm$^{-2}$ according to the Galactic neutral
hydrogen calculator COLDEN from Chandra X-ray center.
This will give an extinction of 0.04 mag at 3.2~\um\
using the extinction cross sections in the Milky Way and SMC provided by
\citet{wei01}\footnote{
Data available at http://www.astro.princeton.edu/~draine/dust/dustmix.html.}.
The results are summarized in Table 1. 
Table 1 shows that the IRC colors of the IR-NE shell is 
(N3/N4, N4/S7, S7/S11)$\sim (0.7, 0.4, 0.8)$. 
For comparison, we have derived  
the fluxes of the bright spot in the IR-NE shell and they 
are listed in Table 1 too. Its IRC colors agree well with those of 
the IR-NE shell. 

The fluxes in Table 1 are the fluxes within ellipses excluding the areas occupied 
by point sources, which amount to a significant portion (15--35\%) of 
the ellipses, 
so that they represent rough estimates for the fluxes of the sources.
The colors, however, should be relatively accurate because they are 
the ratios of fluxes of the same areas. The errors are due to uncertainties in 
the flux calibration and background subtraction.
The absolute flux calibration of the IRC 
array for a point source is accurate to 5--6\% \citep{ona07}. 
The flux calibration for an extended source is different 
from a point source, 
because the incident lights are scattered in the IRC array and 
this diffuse scattered light is not included in the point source 
flux calibration but is partly included for an extended source 
(e.g., \cite{coh07}). According to \citet{sak07}, 4--5\% of the incident light 
of a bright source is lost as scattered light spread in 
horizontal direction in S7 and S11. 
But the correction factor  
depends on the brightness distribution of the source and wavelength, 
and has not been fully studied yet. 
For the IR shells, which are not truly extended sources, 
we consider that 10\% is a reasonable number for the error in color 
associated with this extended-source correction.
Then, including the uncertainties in the background subtraction,
we estimate that the uncertainties in the colors in Table 1 are 
$\simlt 15$\% for the bright spot and $\simlt 30$\% for 
the IR shells.

\begin{table}
  \caption{IRC Fluxes and Colors of the SNR \boone.}\label{tab:first}
  \begin{center}
    \begin{tabular}{lllllllll}
      \hline
            & Area$^a$ & \multicolumn4{c}{Flux} & \multicolumn3{c}{Color} \\
	    & &	 N3  &   N4  &  S7  &  S11 & N3/N4 & N4/S7 & S7/S11 \\
      \hline
IR-NE shell & $84''\times 66''$ &  2.4 &  3.2 & 7.8 &  9.5 &   0.75 &   0.41 &   0.82 \\
Bright spot$^b$ & $30''\times 30''$ &  0.34 &  0.49 & 1.2 &  1.4 &   0.68 &   0.42 &   0.87 \\
IR-SW shell & $43''\times 27''$ &  0.31 &  0.41 & 0.78 &  0.84&   0.75 &   0.53 &   0.93 \\
\hline
RCW 103 (Fast $J$ shock)$^c$ & ...   & ... & ... & ... & ... & 0.01 & 0.30 & 0.65 \\
IC 443 (Molecular shock)$^d$ & ... & ... & ... & ... & ... & 1.1 & 0.44 & 1.1 \\
      \hline
    \end{tabular}
  \end{center}
Note. --- Fluxes are the values (in mJy) at reference wavelengths 
when $F_\nu\propto \nu^{-1}$. The reference wavelengths for 
the N3, N4, S7, and S11 bands are 3.2, 4.1, 7.0, and 11.0~\um.
The fluxes are rough estimates of the total fluxes of the sources. 
The uncertainties in the colors are $\simlt 15$\% for the bright spot
while it is 30\% for the IR shells (see text for details).
For comparison, the expected IRC colors of the Galactic SNRs 
RCW 103 and IC 443 are listed.\hfill\break
$^a$ Major and minor axes of elliptical areas used for the flux calculation.
The areas occupied by point sources are excluded, which amounts to 
15\%, 35\%, and 28\% for the IR-NE shell, Bright spot, and IR-SW shell.\hfill\break
$^b$ The bright spot in the IR-NE shell at 
(01$^{\rm h}$ 06$^{\rm m}$ 24.$^{\rm s}$1, $-72^\circ$ 05$'$ 39$''$). \hfill\break
$^c$ Calculated IRC colors of a bright optical filament in RCW 103  
based on the ISO spectrum of \citet{oli99}. \hfill\break
$^d$ Calculated IRC colors of shocked molecular gas in IC 443 
based on the two temperature model of \citet{rho01}. (See text for 
details.)

\end{table}


The NIR/MIR emission of SNRs could be generally composed of 
ionic forbidden lines, rotational and ro-vibrational 
lines of H$_2$ molecule, PAH (polycyclic aromatic hydrocarbon) bands, 
or continuum emission from hot dust (e.g., see \cite{rea06}).
Synchrotron emission is usually negligible except for young 
Crab-like SNRs. In \eoone, the radio continuum is faint and 
the IR brightness distribution is considered to be different 
from that of radio continuum (\S~3), so that the synchrotron emission 
contribution should be negligible. 
The absence of an associated X-ray emission indicates 
that it is probably not the thermal emission from 
collisionally-heated dust in X-ray emitting hot plasma. 
Also, the ``flat'' IRC color cannot be achieved 
unless the plasma density is unacceptably high 
for usual grain size distribution (see \cite{dra81, dwe86, dwe96}).
The abundance of PAH molecules may be enhanced  
in SNR shells because shocks can convert a 
significant fraction of mass in large grains into 
small grains or PAHs \citep{jon96}. 
But the strong PAH bands are at 
3.3, 6.2, 7.7, 8.6, and 11.3 \um, so that we expect the IRC N4 
band much fainter than the other bands. 
The IRC color of \eoone\ is not consistent with the PAH emission. 
We therefore consider that the observed NIR/MIR emission from 
\eoone\ is either forbidden ionic lines or H$_2$ lines, or both, 
from shock-heated gas.

The existence of the optical filaments associated with the IR shells 
indicates that the IR emission could be at least partly forbidden 
ionic lines and/or H recombination lines from 
fast, ionizing $J$-type shocks 
(hereafter fast $J$ shocks; see \cite{mck84} for 
classification of interstellar shocks) because 
there are optical filaments associated with the IR shells. 
Fig. 3 (top) shows an example of IR spectra 
from fast $J$ shocks dominated by 
ionic lines where we overlay the 
spectral throughput of four IRC bands.
It is an ISO spectrum toward a bright optical 
filament in the Galactic SNR RCW 103.
The strongest lines are [Fe II] 5.34 \um\ (N4), [Ar II] 6.99 \um\ (S7),  
[Ne II] 12.81 \um\ (S11), and [Ne III] 15.56 \um. 
These are also generally the strongest ionic lines 
in this NIR/MIR band (2.0--17.0 \um)  
in other SNRs \citep{dou01a,dou01b,wil06}.
In the band N4, Br$\alpha$ 4.05 \um\ line can be as strong as or even stronger 
than [Fe II] 5.34 \um\ line in principle (e.g., \cite{hol89}). 
But previous NIR observations showed that the ratio of 
Br$\gamma$ 2.166 \um\ to [Fe II] 1.644 \um\ line is usually 
much less ($\sim 1/50$) than theoretical expectations in SNR shocks 
\citep{gra87, oli89, koo07}, which 
suggests that the ratio of Br$\alpha$ to [Fe II] 5.34 \um\ line might 
be much less than the theoretical expectation too.
Note that there are no strong lines in the N3 band (2.6 to 3.8 \um). 
Hydrogen Pf series ($n=8$-11 $\rightarrow$ 5) fall 
in this band, but the total intensity would be 
$\sim 30$\% of Br$\alpha$ line based on the result on 
Case B nebula \citep{hum87}.
Therefore, if the emission is dominated by ionic lines, then 
the flux in N3 should be much weaker than the other three bands. 
The IRC colors of RCW 103, for example, would be 
(N3/N4, N4/S7, S7/S11)$=(0.01, 0.30, 0.65)$
 (Table 1). The observed IRC colors of the IR shells are 
not consistent with a fast $J$ shock dominated by ionic lines
and require H$_2$ molecular lines.

\begin{figure}
  \begin{center}
    \FigureFile(140mm,100mm){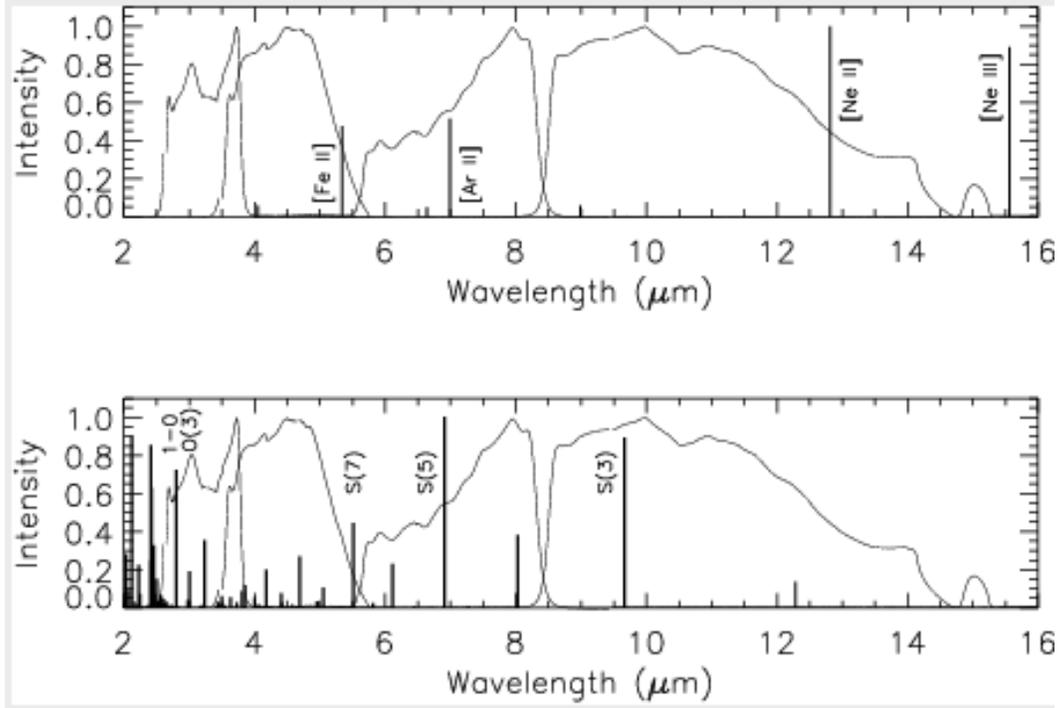}
  \end{center}
  \caption{(top) An ISO spectrum toward a bright optical filament of 
RCW 103 (\cite{oli99}). The throughput of the four IRC bands, 
N3, N4, S7, and S11 from left to right,  
are overlaid. (bottom) A synthesized spectrum of molecular 
shock in IC 443 (see text for an explanation).
}\label{fig:fig3}
\end{figure}

Fig. 3 (bottom) shows a sample spectrum from shocked 
molecular gas dominated by H$_2$ lines. 
It is a synthesized spectrum of IC 443 which is a 
prototypical SNR interacting with molecular clouds. The 
spectrum is based on the empirical, two-temperature
model of \citet{rho01} who showed that the observed 
NIR/MIR spectrum of IC 443 
\citep{ric95, ces99} 
can be fitted well by two components;  
one at $\tex=870$ K with H$_2$ column density 
of $1.3\times 10^{21}$ cm$^{-2}$
and the other at $\tex=2970$ K with $4.0 \times 10^{18}$~cm$^{-2}$. 
The spectrum is dominated by ro-vibrational lines in N3, and 
by pure rotational lines in N4, S7, and S11.  
The expected IRC colors of IC 443 are 
(N3/N4, N4/S7, S7/S11)$=(1.1, 0.44, 1.1)$ (Table 1), which 
are not very different from those of \eoone. 
One might be able to fit the observed color of \eoone\ by a shock 
model where the shock is slow ($\simlt 40$--50~\kms) and 
molecules are not or partially dissociated 
as in IC 443 (e.g., \cite{ric95, ces99}). 
Such numerical studies, however, is beyond the scope of this paper. 
We simply conclude that the observed IR emission is consistent with 
a molecular shock, i.e., a shock propagating into molecular gas, 
although we cannot rule out 
some possible contribution from ionic lines. 

\section{Discussion}

The \akari\ result shows that 
\eoone\ has a SNR shell bright in IR.
According to our analysis,
the IR emission is probably from shocked molecular gas, which 
indicates that \eoone\ is interacting with 
molecular clouds (MCs). 
The centrally-filled thermal X-rays  
support this conclusion because many of the Galactic 
SNRs interacting with MCs do show such X-ray characteristics, 
e.g., \citet{rho98}, \citet{koo03}. 
The faint optical filaments 
associated with the IR shells are not unexpected, because molecular 
clouds are clumpy and shocks propagating through diffuse 
interclump medium could be fast $J$ shocks. Most of 
the observed IR emission, however, might be from slow ($\simlt 40$~\kms) 
non-dissociative, or partially dissociative shocks propagating 
into dense clumps. Such multiple shocks of different natures  
are expected and 
have been observed in several Galactic SNRs interacting 
with molecular clouds (e.g., \cite{che99}, \cite{rea00}, 
\cite{sne05} and references therein). 
A prototypical one is IC 443 (G189.1+3.0; \cite{lee07}). 
IC 443 is interacting with both atomic and molecular gas,
so that, in the optical, its northeastern shell that is 
interacting with diffuse atomic gas is bright, while, in the MIR,  
e.g., in the Midcourse Space Experiment (MSX) 
8.28 \um\ band, the southern shell interacting with molecular gas is bright. 
Its extent (20 pc) is somewhat smaller than \eoone.
It also has faint optical filaments over the entire remnant.
X-rays are bright inside the remnant, which is of 
thermal origin ($\sim 1$ keV; \cite{kaw02}).
Therefore, \eoone\ is quite similar to IC 443 and 
might be a
middle-aged ($\simgt 10^4$~yrs) SNR expanding in a multiphase 
interstellar medium.

The interaction with MCs suggests that \eoone\ is 
possibly a core-collapse SNR with a massive progenitor star. 
Our analysis indicates that the radio nebula is 
probably inside the IR shell, and it could be 
a pulsar wind nebula (PWN) considering its flat spectrum.  
The radio nebula appears too large for a 
middle-aged PWN, but it is possible that the nebula can be 
rejuvenated through the interaction with 
the SNR shell as in the Galactic SNR CTB 80 \citep{fes88, koo90}. 
CTB 80 is another SNR where the SNR shell is detected in IR, 
i.e., in an IRAS 60 to 100 \um\ color map. 
It is an old ($\sim 1\times 10^5$~yr) 
SNR where apparently its pulsar is interacting with the SNR shell 
to produce a large ($\sim 40$ pc) synchrotron nebula. 
The flux of \eoone\ ($\sim 13$ mJy at 1 GHz) is 
an order of magnitude less than that of CTB 80, which 
would be $\sim 130$~mJy if it were in the SMC.

On the other hand, there have been suggestions that 
\eoone\ is a Type Ia SNR \citep{hug94, hey04}.
It is not impossible for a Type Ia SNR to be interacting 
with MCs, e.g., Tycho \citep{lee04}. 
The radio nebula may have a morphology similar 
to the optical image rather than the IR image 
as in IC 443, although its spectral index is 
rather unusual for a shell-type SNR. 
A multiwavelength study with high-resolution radio observations 
is needed to resolve the issue.

\section{Conclusion}

\eoone\ is the first SNR detected in the 2.6--15 \um\ bands in the SMC. 
It is the faintest known SMC SNR in the radio and 
is not particularly bright in X-rays. 
The brightest two SMC SNRs in X-rays are 
SNRs B0102$-$72.3 (IKT 22) and B0103$-$72.6 (IKT 23),
which are both young oxygen-rich SNRs \citep{hey04}.  
B0102$-$72.3 is also one of the brightest radio SNRs.
B0102$-$72.3 was observed by 
\citet{sta05} using the {\it Spitzer Space Telescope} and 
was detected at 24 \um\ (80 mJy), but neither at 8 nor 70 \um, 
which was interpreted as thermal emission from hot dust 
at $\sim 120$ K.
The $1\sigma$ upper limit at 8 \um\ was 3 mJy.  
B0103$-$72.6 was observed by us using \akari\ 
(see \S~2), but 
no NIR/MIR emission associated with the SNR shell 
was detected. The discovery of \eoone\ 
shows that IR bright SNRs are not necessarily 
radio or X-ray bright ones, or vice versa.

\eoone\ is bright in the IR probably because it is 
interacting with MCs.
If \eoone\ were in our Galaxy, say at a distance of 1.5 kpc, its flux would be 
$\sim 16$~Jy in the S7/S11 bands. For comparison, 
the flux of IC 443, which is at the same distance, 
estimated from the MSX data is $\sim 50$ Jy at 8.28 \um.
Therefore, \eoone\ would 
be a bright, but not an exceptionally bright SNR in the Galaxy.
It is also significantly fainter than 
the brightest NIR/MIR SNRs in the 
Large Magellanic Cloud, e.g., 
N49 and N63A, whose fluxes are 32--41 mJy and 
130--300 mJy in the {\it Spitzer} IRAC 3.6/4.5 and 
5.8/8.0 \um\ bands, respectively (\cite{wil06}).
N49 is also known to be interacting with MCs \citep{ban97}.
Whether or not \eoone\ is the brightest NIR/MIR SNR in the SMC 
needs to be answered.

Our work shows that 
the \akari\ IRC colors can be used for shock diagnosis.
In particular, the brightness of 
N3 relative to the other bands is a good 
indicator that distinguishes between ionic and molecular shocks. 
A careful analysis of IRC colors might enable us to 
constrain the shock parameters. 

We would like to thank all the members of the \akari\ project
for their dedicated work. 
\akari\ is a JAXA project with the participation of ESA.
This work was supported by the Korea Research 
Foundation (grant No. R14-2002-058-01003-0).




\begin{thebibliography}{}
\bibitem[Banas et al.(1997)]{ban97} Banas, K. R., Hughes, J. P., Bronfman, L., \& 
Nyman, L.-\AA\ 1997, ApJ, 480, 607
\bibitem[Cesarsky et al.(1999)]{ces99} Cesarsky, D., Cox, P., Pineau des 
Forets, G., van Dishock, E. F., Boulanger, F., \& Wright, C. M.\ 1999, 
A\&A, 348, 945
\bibitem[Cohen et al.(2007)]{coh07} Cohen, M. et al.\ 2007, MNRAS, 374, 979
\bibitem[Chevalier(1999)]{che99} Chevalier, R. A.\ 1999, ApJ, 511, 798
\bibitem[Draine(1981)]{dra81} Draine, B. T.\ 1981, ApJ, 245, 880
\bibitem[Douvion et al.(2001a)]{dou01a} Douvion, T., Lagage, P. O., Cesarsky, C. J., Dwek, E.\ 2001, A\&A, 373, 28
\bibitem[Douvion, Lagage, \& Pantin(2001b)]{dou01b} Douvion, T., 
Lagage, P. O., Pantin, E.\ 2001, A\&A, 369, 589
\bibitem[Dwek(1986)]{dwe86} Dwek, E.\ 1986, ApJ, 302, 363
\bibitem[Dwek, Foster, \& Olaf(1996)]{dwe96} Dwek, E., Foster, S. M., \& Olaf, V.\ 1996, ApJ, 457, 244
\bibitem[Fesen, Shull, \& Saken(1988)]{fes88} Fesen, R. A., Shull, J. M., Saken, J. M.\ 1988, Nature, 334, 229
\bibitem[Filipovi\'c et al.(2005)]{fil05} 
Filipovi\'c, M. D., Payne, J. L., Reid, W., Danforth, C. W., 
Staveley-Smith, L., Jones, P. A., White, G. L.\ 2005, MNRAS, 364, 217
\bibitem[Graham, Wright, \& Longmore(1987)]{gra87}
Graham, J. R., Wright, G. S., \& Longmore, A. J.\ 1987, ApJ, 313, 847
\bibitem[Hollenbach \& McKee(1989)]{hol89} Hollenbach, D. \& McKee, C. F.\ 1989, 
ApJ, 342, 306
\bibitem[Hughes et al.(1994)]{hug94} Hughes, J. P., Smith, R. C.\ 1994, AJ, 107, 1363
\bibitem[Hummer \& Storey(1987)]{hum87} Hummer, D. G. \& Storey, P. J.\ 1987, MNRAS, 224, 801
\bibitem[Ita \& Pearson(2007)]{ita07} Ita, Y. \& Peason, C.\ 2007, 
IRC Imaging Data Reduction Pipeline User Manual
\bibitem[Jones, Tielens, \& Hollenbach(1996)]{jon96} Jones, A. P., Tielens, A. G. G. M., 
\& Hollenbach, D. J.\ 1996, ApJ, 469, 740
\bibitem[Kawasaki et al.(2002)]{kaw02} Kawasaki, M., Ozaki, M., Nagase, F., Masai, K., 
Ishida, M., \& Petre, R.\ 2002, ApJ, 572, 897
\bibitem[Koo(2003)]{koo03} Koo, B.-C.\ 2003, in 
The Proceedings of the IAU 8th Asian-Pacific Regional Meeting 
(ASP Conference Proceedings v. 289), ed Ikeuchi, S., Hearnshaw, J. \& 
Hanawa, T.\ (San Francisco: ASP), page 199
\bibitem[Koo et al.(1990)]{koo90} Koo, B.-C., Reach, W. T., Heiles, C., Fesen, R. A., \& 
Shull, J. M.\ 1990, ApJ, 364, 178
\bibitem[Koo et al.(2007)]{koo07} Koo, B.-C., Moon, D.-S., Lee, H.-G., Lee, J.-J., 
\& Mathews, K.\ 2007, ApJ, 657, 308
\bibitem[Lee et al.(2004)]{lee04} Lee, J.-J., Koo, B.-C., \& Tatematsu, K.\ 2004, ApJL,
605, 113
\bibitem[Lee et al.(2007)]{lee07} Lee, J.-J., Koo, B.-C., Yun, M. S., 
Stanimirovi\'c, S., Heiles, C., Heyer, M.\ 2007, AJ, submitted
\bibitem[Mathewson et al.(1984)]{mat84}
   Mathewson, D. S., Ford, V. L., Dopita, M. A., Tuohy, I. R., Mills, B. Y., 
Turtle, A. J.\ 1984, ApJS, 55, 189
\bibitem[McKee et al.(1984)]{mck84} McKee, C. F., Chernoff, D. F., \& Hollenbach, 
D. J.\  1984, in Galactic and Extragalactic Infrared Spectroscopy, ed. Kessler, M. F.
\& Phillips, J. P.\ (Dordrecht: D. Reidel), page 103
\bibitem[Oliva, Moorwood, \& Danziger(1989)]{oli89} Oliva, E., Moorwood, A. F. M., \& Danziger, I. J.\ 1989, A\&A, 214, 307
\bibitem[Oliva et al.(1999)]{oli99} Oliva, E., Moorwood, A. F. M., Drapatz, S., Lutz, D., \& Sturm, E.\ 1999, A\&A, 343, 943
\bibitem[Onaka et al.(2007)]{ona07} Onaka, T. et al.\ 2007, in this volume
\bibitem[Park et al.(2003)]{park03} Park, S., Hughes, J. P., Burrows, D. N., Slane, P. O., 
Nousek, J. A., \& Garmire, G. P.\ 2003, ApJL, 598, L95
\bibitem[Reach \& Rho(2000)]{rea00} Reach, W. T., \& Rho, J.\ 2000, ApJ, 544, 843
\bibitem[Reach et al.(2006)]{rea06} Reach, W. T. et al.\ 2006, AJ, 131, 1479
\bibitem[Rho \& Petre(1998)]{rho98} Rho, J. \& Petre, R.\ 1998, ApJL, 503, L167	
\bibitem[Rho et al.(2001)]{rho01} Rho, J., Jarrett, T. H., Cutri, R. M., \& Reach, W. T.\ 2001, ApJ, 547, 885
\bibitem[Richter, Graham, \& Wright(1995)]{ric95} Richter, M. J., 
Graham, J. R., \& Wright, G. S.\ 1995, ApJ, 454, 277
\bibitem[Sakon et al.(2007)]{sak07} Sakon, I. et al.\ 2007, in this volumne
\bibitem[Snell et al.(2005)]{sne05} Snell, R. L., Hollenbach, D., Howe, J. E., 
Neufeld, D. A., Kaufman, M. J., Melnick, G. J., Bergin, E. A., \& Wang, Z.\ 2005, 
ApJ, 620, 758
\bibitem[Stanimirovi\'c et al.(2005)]{sta05} 
Staminirovic\', S., Bolatto, A. D., Sandstrom, K., Leroy, A. K., Simon, J. D., 
Gaensler, B. M., Shah, R. Y., Jackson, J. M.\ 2005, ApJ, 632, L103
\bibitem[van der Heyden et al.(2004)]{hey04}
van der Heyden, K. J., Bleeker, J. A. M., Kaastra, J. S.\ 2004, A\&A, 421, 1031
\bibitem[Weingartner \& Draine(2001)]{wei01} Weingartner, J.C. \& Draine, B.T.\ 2001, 
ApJ, 548, 296
\bibitem[Williams, Chu, \& Guendl(2006)]{wil06}Williams, R. M., Chu, Y.-H., \& Gruendl, R.\ 2006, ApJ, 132,1877

\end{thebibliography}
\end{document}